# Superconducting weak links at substrate steps in bilayer nickelate films

Yidi Liu[1,2†], Yaoju Tarn[1,3†], Lopa Bhatt[4], Puhua Zhang[5], Joonseo Song[1,2], Berit H. Goodge[6], David A. Muller[4,7], Harold Y. Hwang[1,3], Yijun Yu[1,3]*

[1] Stanford Institute for Materials and Energy Sciences, SLAC National Accelerator Laboratory, Menlo Park CA 94025, USA
[2] Department of Physics, Stanford University, Stanford CA 94305, USA
[3] Department of Applied Physics, Stanford University, Stanford CA 94305, USA
[4] School of Applied and Engineering Physics, Cornell University, Ithaca NY 14853, USA
[5] State Key Laboratory of Surface Physics and Department of Physics, Fudan University, Shanghai 200433, China
[6] Max Planck Institute for Chemical Physics of Solids, 01187 Dresden, Germany
[7] Kavli Institute at Cornell for Nanoscale Science, Cornell University, Ithaca, NY 14853, USA

[†]Equal contribution
*yuyijun@stanford.edu

Keywords: bilayer nickelate, thin film, superconductivity, superconducting weak links

**Abstract:** Ambient-pressure superconductivity above 40 K has been realized in bilayer nickelate thin films on $SrLaAlO_4$ (001), yet resistive measurements often exhibit incomplete zero-resistance states and/or two-step superconducting transitions of unresolved origin. We show that these features arise from superconducting weak links at substrate step-terrace structures. A direct correlation between the azimuthal-angle-dependent two-step transition and step-terrace orientation establishes this connection, further supported by atomic-scale visualization of structurally-disrupted regions at step terraces. Modeling the films as pristine (weak-link) regions with higher (lower) transition temperature reproduces essential features of the two-step transition using both a simple model and finite-element simulations. Transport measurements further reveal a strong dependence of transition characteristics — and even the occurrence of superconductivity itself — on average terrace width. These results identify step-induced weak links as the microscopic origin of two-step transitions in bilayer nickelate films, providing a framework for understanding anisotropic and percolative transport, and a route to superconducting devices.

The realization of superconductivity in bilayer nickelate thin films under epitaxial compressive strain[1–4] provides a promising route to study nickelate superconductivity without the requirement of high pressure[5–7]. However, the thin-film approach also introduces additional structural and chemical complexities that can strongly influence superconducting properties. Although compressive strain is required to stabilize superconductivity, achieving robust superconductivity in these films remains challenging due to the large lattice mismatch between the nickelates and $SrLaAlO_4$ (001) (SLAO) substrates, the metastable nature of the bilayer phase[8], and the strong sensitivity of superconductivity to oxygen off-stoichiometry[9]. Since the first report of superconductivity in bilayer nickelate thin films, a characteristic two-step superconducting transition has been repeatedly observed across different compositions and growth conditions[1–4,10–22] (**Figure 1a**). Despite its ubiquity, the physical origin of this two-step transition remains unresolved, with proposed explanations ranging from sample inhomogeneity[1,2,20] to more exotic scenarios such as spin-glass–like states in this newly discovered superconducting system[13]. More broadly, two-step superconducting transitions have been reported in a variety of systems, including 1/8 Ba-doped cuprates, where they are linked with spin ordering[23], and iron-based superconductors, where they are associated with phase separation[24], both of which can induce Josephson-coupling effects similar to those observed in ordered[25,26] and disordered[27,28] Josephson-coupled circuits. These observations raise a key question: is the two-step transition an intrinsic manifestation of an underlying electronic instability of bilayer nickelate films, or is it induced by disorder, and if so, by what mechanism? In this work, we investigate the physical origin of the two-step transition and reveal the critical role of substrate terraces in determining the superconducting properties of bilayer nickelate thin films.

**Figure 1** presents temperature-dependent resistivity, normalized to its value at 60 K ($\rho(T)/\rho(60$ K)), for representative superconducting transitions in bilayer nickelate thin films measured using two orthogonal van der Pauw (vdP) configurations, with sample edges aligned along the pseudo-cubic axes of the substrate. The two-step transition is observed in multiple bilayer nickelate compositions, including $La_2PrNi_2O_7$ and $La_2SmNi_2O_7$ (**Figure 1b**–**d** and **Figure 1e**–**g**, respectively). Despite this apparent universality, the detailed characteristics of the two-step transition exhibit pronounced sample-to-sample variations. In some samples, the two vdP configurations yield nearly identical resistivity evolutions (**Figure 1b**,**e**), whereas in others, pronounced anisotropy emerges: one configuration displays a single-step transition while the orthogonal configuration retains a clear two-step feature (**Figure 1c**,**f**). In the most extreme cases, the lower-temperature transition dominates the transport response in one configuration (**Figure 1d**,**g**).

The coexistence of robustness and apparent randomness in these observations is particularly intriguing. While the two-step transition persists across different compounds, the detailed transport anisotropy varies significantly from sample to sample and shows no systematic correlation with the crystallographic axes. Although random disorder could account for the absence of directional preference, it does not readily explain the reproducibility and persistence of a distinct two-step superconducting transition as opposed to a continuously broadened transition.

To gain insight into the origin of the two-step transition, we fabricated a thin-film device with electrodes arranged in a circular geometry to probe the azimuthal-angle dependence of the superconducting transition (**Figure 2a**). Topographic atomic force microscopy (AFM) and its two-dimensional correlation map reveals a film surface with a step-terrace orientation running from the lower left to the upper right (**Figure 2b**, **c**). As shown in **Figure 2d**, the color-coded transport data from different vdP configurations reveal a clear correlation with the step-terrace direction: when the current flows predominantly parallel to the step edges (purple configuration), a single-step superconducting transition is observed, whereas a pronounced two-step transition emerges when the current flows perpendicular to the step terraces (red configuration). The remaining configurations exhibit intermediate behavior between these two limits. This behavior is further visualized in the polar plot of the resistivity at 20 K, normalized to its value at 60 K, as a function of current azimuthal angle (**Figure 2e**). The normalized resistivity approaches zero when the current flows parallel to the step-terrace direction and reaches a maximum when the current is oriented perpendicular to the step terraces.

The clear correlation between the step-terrace direction and the measured resistivity suggests that the microscopic structure at these step terraces plays a decisive role. In oxide substrates, step terraces can act as extended structural defects, particularly when the substrate step height is incompatible with the structural periodicity of the epitaxial film. In the case of SLAO (001), the step height (6.3 Å) does not match the thickness of a bilayer nickelate unit (10.3 Å), making the formation of line defects at step terraces unavoidable during the initial epitaxial growth. As shown in **Figure 2g**, annular dark-field scanning transmission electron microscopy (ADF-STEM) directly reveals structurally disrupted regions separating adjacent crystalline domains in a bilayer nickelate film (also seen in ref. [29]). Such disrupted regions can naturally act as superconducting weak links between neighboring terraces (as schematically illustrated in **Figure 2f** and **Figure S1**), giving rise to a strong directional dependence of the superconducting transport. A similar concept of superconducting weak links associated with substrate steps have also been reported in ultrathin $YBa_2Cu_3O_7$ films grown on $SrTiO_3$ (ref. [30]), albeit with a much weaker effect. This comparison nevertheless supports the physical plausibility of step-terrace–induced weak-link behavior in superconducting bilayer nickelates thin films. We note that occasionally the step height becomes locally compensated by the formation of secondary $(La,Pr)_2NiO_4$ phases (**Figure S2a**), similar to observations reported for $Sr_2RuO_4$ films[31], Nevertheless, structural analysis still reveals residual strain effects (**Figure S2b**,**c**), which likely generate weak links given the strong sensitivity of superconductivity to local structural distortions[29]. We note that both mechanisms can occur – weak links induced by step-triggered extended defects (**Figure S1**), and weak links induced by residual strain relaxation around otherwise coherently accommodated steps (**Figure S2**) – and we cannot quantitatively resolve which mechanism is dominant.

To validate the proposed step-terrace–induced weak-link scenario, we construct a minimal model that captures the essential ingredients of the observed transport behavior. We consider a sample composed of two distinct regions: region A represents pristine bilayer nickelate with a superconducting transition temperature ~ 40 K, while region B corresponds to step-terrace weak-link regions whose $\rho(T)$ reaches zero only at a lower temperature, reflecting the temperature scale at which superconducting coherence is established cross the weak-link regions. We also validate that the lateral size of the strain-disrupted region $L_{\text{strain}}$ ~ 1-5 nm (i.e. the barrier thickness, estimated from **Figure S1b,c** and **S2b,c**) is of the same order of magnitude as the Ginzburg-Landau in-plane coherence length $\xi_{ab}$ ~ 2-3 nm of bilayer nickelate thin films grown on SLAO[1], indicating that individual step terraces function in a regime of Josephson-type weak links. The assumed $\rho(T)$ of the two regions are shown schematically in **Figure 3a**.

When region B is oriented along the diagonal of the square sample, the two vdP configurations yield nearly identical $\rho(T)$ curves, as illustrated in **Figure 3b**. In contrast, when region B is aligned parallel to one edge of the square sample, the two vdP configurations probe the weak-link region differently, resulting in a single-step transition in one configuration and a two-step transition in the orthogonal configuration (**Figure 3c**). As the fractional area of region B increases, or equivalently as its resistivity contribution becomes larger, the configuration exhibiting a two-step transition evolves toward behavior increasingly dominated by region B, closely mimicking the $\rho(T)$ for region B alone (**Figure 3d**). **Figure 3b-d** therefore captures the key transport features observed in **Figure 1**.

To provide a more realistic comparison with experiment, we further perform finite-element simulations of configuration-dependent $\rho(T)$, shown in **Figure 4**. **Figure 4a** presents the azimuthal-angle dependence of $\rho(T)$ measured in an approximately in-line configuration, while **Figure 4c** shows $\rho(T)$ obtained for different electrode separations with current flowing parallel or perpendicular to the step-terrace direction. The simulations are performed using the geometry illustrated in **Figure S3a**, where stripe patterns represent alternating regions A and B with their corresponding $\rho(T)$ defined in **Figure S3b**. The close qualitative agreement between the experimental data (**Figure 4a**,**c**) and the simulations (**Figure 4b**,**d**) supports the validity of our weak-link–based interpretation. The magnitude of the lower-$T_c$ transition is strongly modulated by the measurement geometry, providing a natural explanation for the wide variations observed in **Figure 1** and reported across the literature[1–4,10–15,17,19–21]. To simulate the specific cases shown in **Figure 1c**,**f**, we further adjust the simulation parameters to reflect stronger weak-link contributions, as presented in **Figure S4**.

Based on this scenario, we now discuss the impact of miscut angle or terrace width of SLAO (001) substrate. Experimentally, we frequently observe that films grown on substrates with larger step-terrace widths (corresponding to smaller miscut angles) more readily reach zero resistance. As shown in **Figure 5**, the sample with a nominal average step-terrace width of approximately 920 nm and 540 nm reaches zero resistance (**Figure 5a**,**b**,**e**,**f**), whereas the sample with an averaged terrace width of approximately 250 nm does not (**Figure 5c**,**g**). For the sample with an averaged terrace width of 120 nm, superconductivity is completely absent (**Figure 5d**,**h**), despite exhibiting comparable or even better crystallinity as indicated by X-ray diffraction (XRD, **Figure S5**) and undergoing the same post-growth ozone anneal protocol[3]. This behavior cannot be simply understood by comparing the idealized geometries shown in **Figure 3c**,**d**, where both measurement configurations ultimately reach zero resistance. Instead, it is essential to recognize that real step terraces on SLAO differ substantially from the straight boundaries assumed in our toy model or simulation.

In practice, step edges on SLAO (001) are often irregular and meandering (**Figure S6**), reflecting the absence of a strong crystallographic locking direction and the tendency of the surface to minimize step energy through local roughening. Consequently, narrow terraces (larger miscut angles) form a dense and fragmented network of weak-link segments, in which cross-talk between neighboring step edges more effectively disrupts global superconducting percolation and destroy the zero-resistivity state. In the case of **Figure 5d**,**h**, we suspect that the fragmented network appears sufficiently dense that superconductivity is suppressed not only at the mesoscopic percolation level but even locally within individual patches or islands. Although the nominal average step-terrace width (~100 nm) remains much larger than the superconducting coherence length (on the order of several nanometers[1]), STEM measurements reveal frequent interfacial roughening on a much shorter length scale (~10 nm), arising from the meandering of step edges. Given that the local atomic (**Figure S1**,**S2**) and electronic structure near terrace boundaries can be strongly modified by strain mismatch[29], together with the expectation of enhanced strain relaxation in regions with densely spaced step edges, the effective critical length scale required to sustain superconductivity is likely substantially larger than the coherence length alone, with additional suppression potentially arising from inverse proximity effects[32]. Within this framework, the complete suppression of superconducting fluctuations becomes plausible. A comparable domain length scale has also been reported in bulk infinite-layer nickelates[33], which do not exhibit superconductivity, suggesting a potentially common underlying origin.

These findings also rationalize the low critical current densities reported even in the best bilayer nickelate thin films[3,4], as superconducting weak links act as the limiting factor for global current flow. Moreover, this picture naturally explains why the superconducting onset temperature inferred from mutual inductance measurements is often lower than that determined from transport[3,15]. When the current is directed parallel to the step-edge direction, zero resistance can be achieved following a single-step transition in transport measurements. In contrast, mutual inductance probes global diamagnetic screening, which can only develop after superconducting coherence is established across the weak-link network, corresponding to the lower-temperature transition. Notably, such two-step behavior is not observed in the superconducting transition of bilayer nickelate films grown on $LaAlO_3$ substrates at ambient[34] or high pressure[35], or $NdGaO_3$ substrates under pressure[35], where the smaller step height, reduced strain, larger film thickness, and/or lower superconducting transition temperature likely act together to suppress the formation or impact of step-terrace–induced weak links.

We further note that in $YBa_2Cu_3O_7$ thin films, step-terrace–induced weak-link effects are observed primarily in the ultrathin limit[30] and the Cu-O planes recover structural continuity as the film thickness increases beyond approximately 7 bilayers [36,37]. In contrast, bilayer nickelate films grown on SLAO face competing constraints: increasing the film thickness is desirable to mitigate weak-link effects associated with step terraces, yet also leads to strain relaxation that suppresses superconductivity or promotes the formation of competing phases[11]. Balancing these competing requirements—maintaining sufficient thickness to reduce weak-link connectivity while preserving the epitaxial strain necessary for superconductivity—remains a key challenge for future experiments.

In summary, we identify step-terrace–induced superconducting weak links as the microscopic origin of the widely observed two-step superconducting transition in bilayer nickelate thin films grown on SLAO. By combining angle-dependent transport measurements, simple model analysis, finite-element simulations, and atomic-scale STEM imaging, we show that the interplay between the step-terrace geometry and weak-link percolation governs the anisotropic and configuration-dependent superconducting transport. This framework reconciles a wide range of experimental observations in the literature, rationalizes the probe-dependent superconducting temperatures, and highlights substrate morphology as a key control parameter. Our results establish step-terrace engineering as an essential consideration for optimizing superconductivity in metastable bilayer nickelates and, more broadly, in ultrathin superconducting films.

**Supporting Information**

Structural analysis of a disrupted film region near a step terrace; Structural accommodation near a step terrace; Simulation details; Symmetric XRD $\theta$-$2\theta$ scans of the four samples in **Figure 5**; AFM topography of a representative as-received SLAO substrate.

**Acknowledgements**

The authors thank Y. Lee and B. Y. Wang for helpful discussions, and E. K. Ko for early-stage observations. This work was support by the US Department of Energy, Office of Basic Energy Sciences, Division of Materials Sciences and Engineering (DE-AC02-76SF00515). Work at the Stanford Nano Shared Facilities (SNSF) RRID:SCR_023230 is supported by the National Science Foundation (ECCS-1542152). L.B. and D.A.M. acknowledge support by the NSF Platform for the Accelerated Realization, Analysis, and Discovery of Interface Materials (PARADIM) under cooperative agreement No. DMR-2039380. B.H.G. was supported by the Max Planck Society. The Thermo Fisher Spectra 300 X-CFEG was acquired with support from PARADIM, an NSF-MIP (DMR-2039380), and Cornell University. STEM related work made use of the Cornell Center for Materials Research shared instrumentation facility. Y.T. acknowledges support from the Stanford Graduate Fellowship (SGF).

**Methods**

*Sample preparation, characterization, and transport measurements*

Thin films of $La_2PrNi_2O_7$ and $La_2SmNi_2O_7$ were grown on $SrLaAlO_4$ (001) substrates by pulsed laser deposition, characterized by XRD diffraction, and underwent post-growth ozone annealing and transport measurements following the methods detailed in refs. [3,18].

*STEM measurements*

To prepare cross-sectional lamellas, we used a standard focused ion beam (FIB) lift-out procedure using a Thermo Fisher Helios G4 UX FIB. We performed STEM imaging in ADF mode using a Cs-corrected Thermo Fisher Scientific (TFS) Spectra 300 X-CFEG 300 kV with a probe convergence semi-angle of 30 mrad. Inner/outer collection angles in ADF geometry were 77/200. Lattice rotation and spacing measurements were performed using KEMSTEM-package described in ref. [38].

**Author Contributions**

Y.L. and Y.T. contributed equally to this work. Y.Y. and H.Y.H. supervised and conceived the project. Y.L. and Y.T. grew and characterized the samples. Y.Y., Y.L., and Y.T. fabricated devices and measured transport properties, with help from J.S. L.B., B.H.G., and D.A.M. conducted the STEM measurements. Y.Y., Y.L., and Y.T. analyzed the data. P.Z. performed numerical simulations. Y.Y. wrote the manuscript with input from all authors.

**Data Availability Statement**

Source data is available upon reasonable requests to the first and/or corresponding authors.

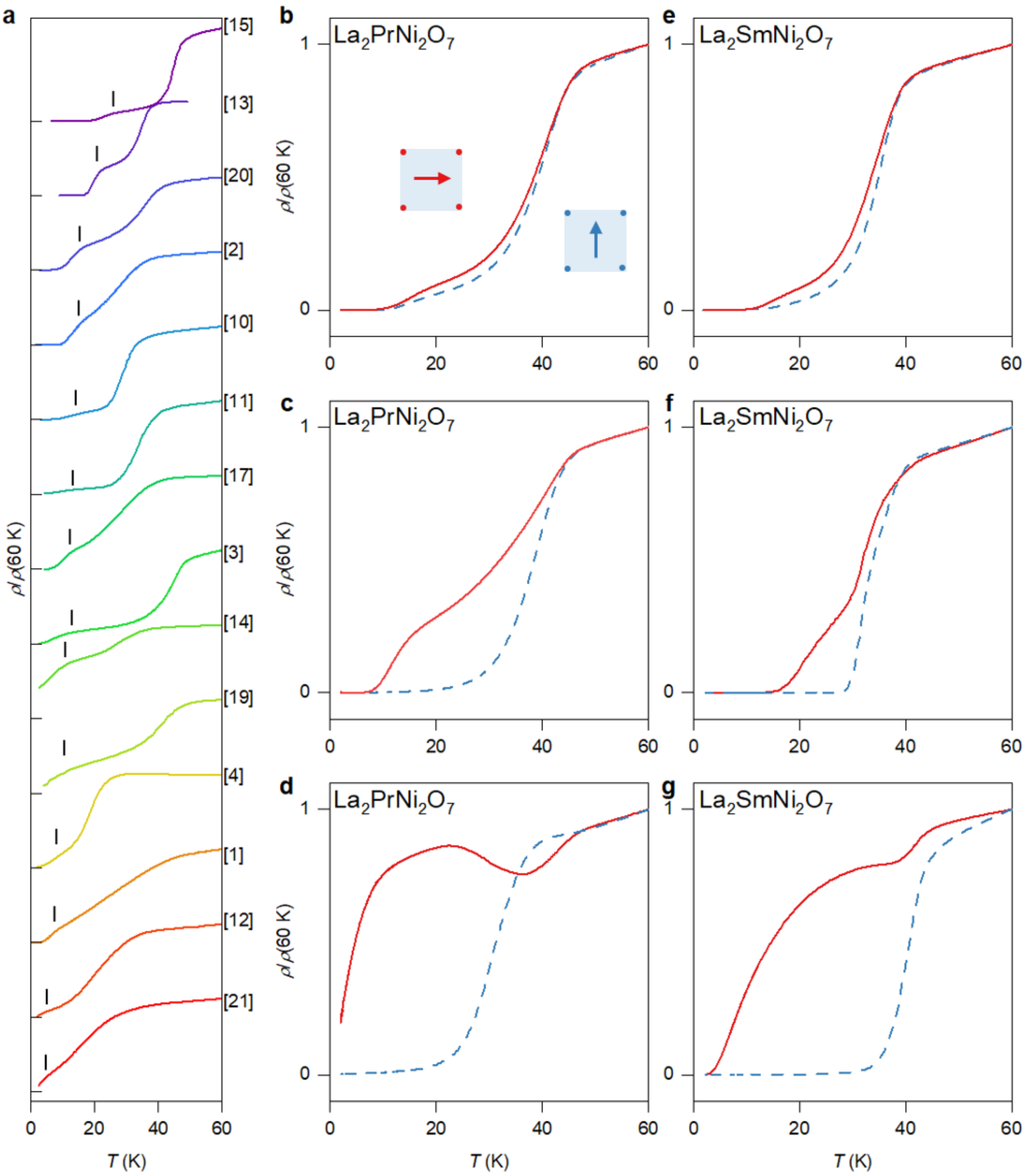

**Figure 1. Two-step superconducting transitions in bilayer nickelates thin films grown on SLAO. (a)** Two-step superconducting transitions reported in the literature[1–4,10–15,17,19–21]. $\rho(T)/\rho(60\ \mathrm{K})$ are arranged from bottom to top in approximate order of increasing lower-transition-step temperature, indicated by the vertical short lines. Here, we deliberately replot, for each reference, the dataset exhibiting the most pronounced two-step behavior. Transport measurements in the van der Pauw geometry for **(b-d)** $La_2PrNi_2O_7$; **(e-g)** $La_2SmNi_2O_7$. The samples are approximately square in shape, with their edges aligned along the pseudo-cubic axes of the substrate. Electrical contacts were deposited at the four corners, and transport measurements were performed using two van der Pauw configurations, in which current was applied between adjacent electrodes while the voltage was measured across the remaining pair, as illustrated in the inset of **(b)**. **(b)** and **(e)** show representative samples in which the superconducting transitions measured in the two configurations are nearly identical but with signatures of two-step–like transitions. **(c)** and **(f)** display samples with pronounced anisotropy, where one configuration exhibits a single sharp transition while the other shows an obvious two-step–like transition. **(d)** and **(g)** represent the extreme limit of this behavior: in one configuration (red curve), the lower-temperature transition dominates the transport response, whereas in the other configuration (blue curve), the higher-temperature transition is dominant.

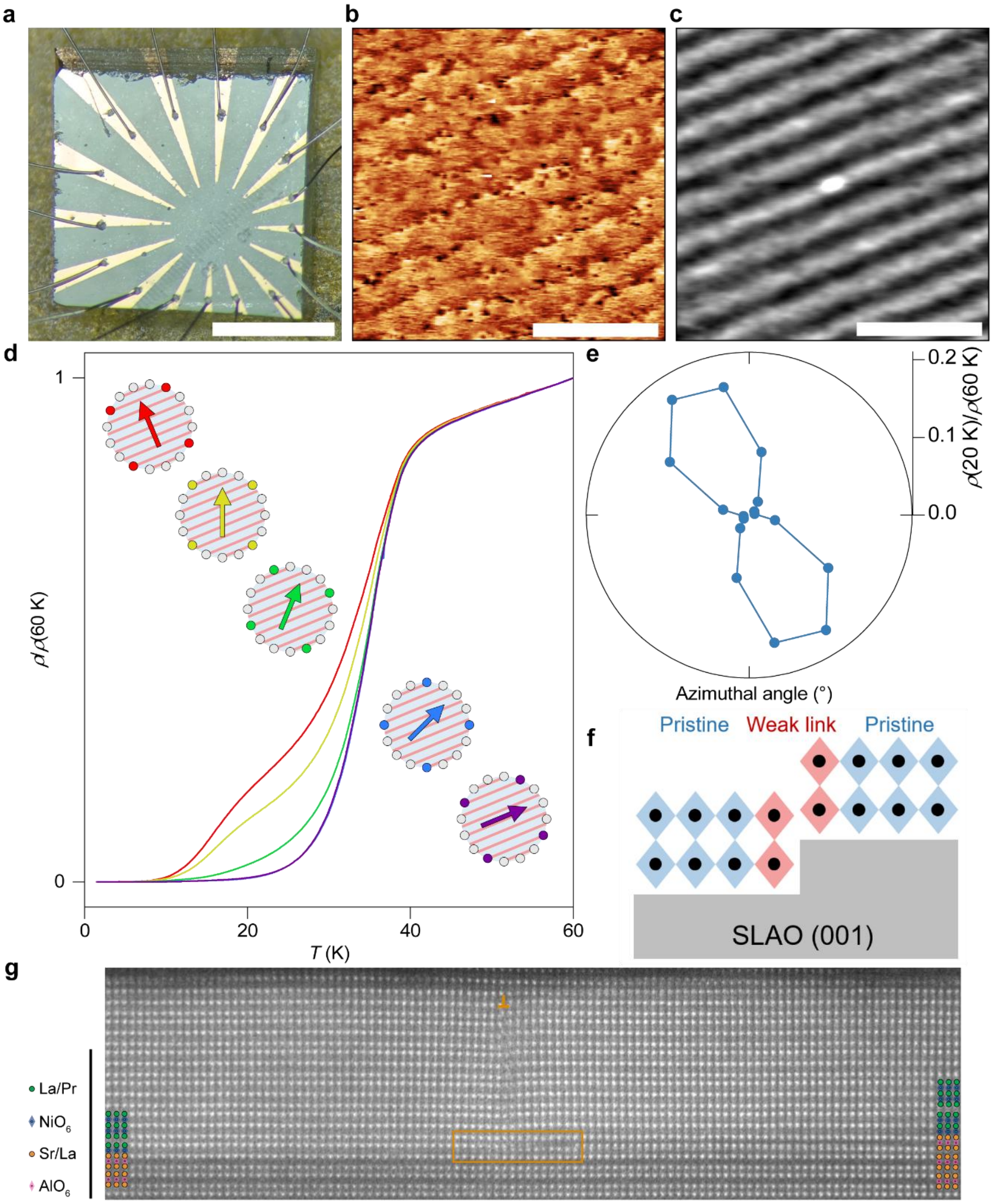

**Figure 2. Azimuthal-angle dependence of superconducting transitions. (a)** Optical image of a $La_2SmNi_2O_7$ thin-film device patterned with electrodes arranged in a circular geometry, enabling transport measurements along multiple in-plane directions. **(b)** AFM topography of the same sample in **(a)**, revealing the step-terrace morphology. The contour spans a height range of 1 nm. **(c)** The corresponding two-dimensional autocorrelation map, which highlights the orientation of the terraces. **(d)** Temperature-dependent normalized resistivity measured for different azimuthal current directions, demonstrating a pronounced angular dependence of the superconducting transition. Insets illustrate the color-coded measurement configurations: arrows indicate the current direction, and colored dots mark the active electrode pairs. The stripe orientation in each inset corresponds to the step-terrace direction extracted from **c**. Note that the blue curve overlaps with the purple curve. **(e)** Azimuthal-angle dependence of the resistivity at 20 K, normalized to its value at 60 K, extracted from the data in **(d)**. **(f)** Schematic illustration

of the pristine region and the weak-link region of the sample along a substrate step terrace. **(g)** ADF-STEM image showing two laterally adjacent crystalline domains separated by a narrow region with misfit atomic structures at the step terrace. The interfacial region highlighted by the yellow box exhibits disrupted lattice registry, suggesting a structurally distinct weak-link region between the neighboring domains. Scale bar in **(a)**: 1 mm; **(b)**, **(c)**: 2 μm; **(g)**: 5 nm.

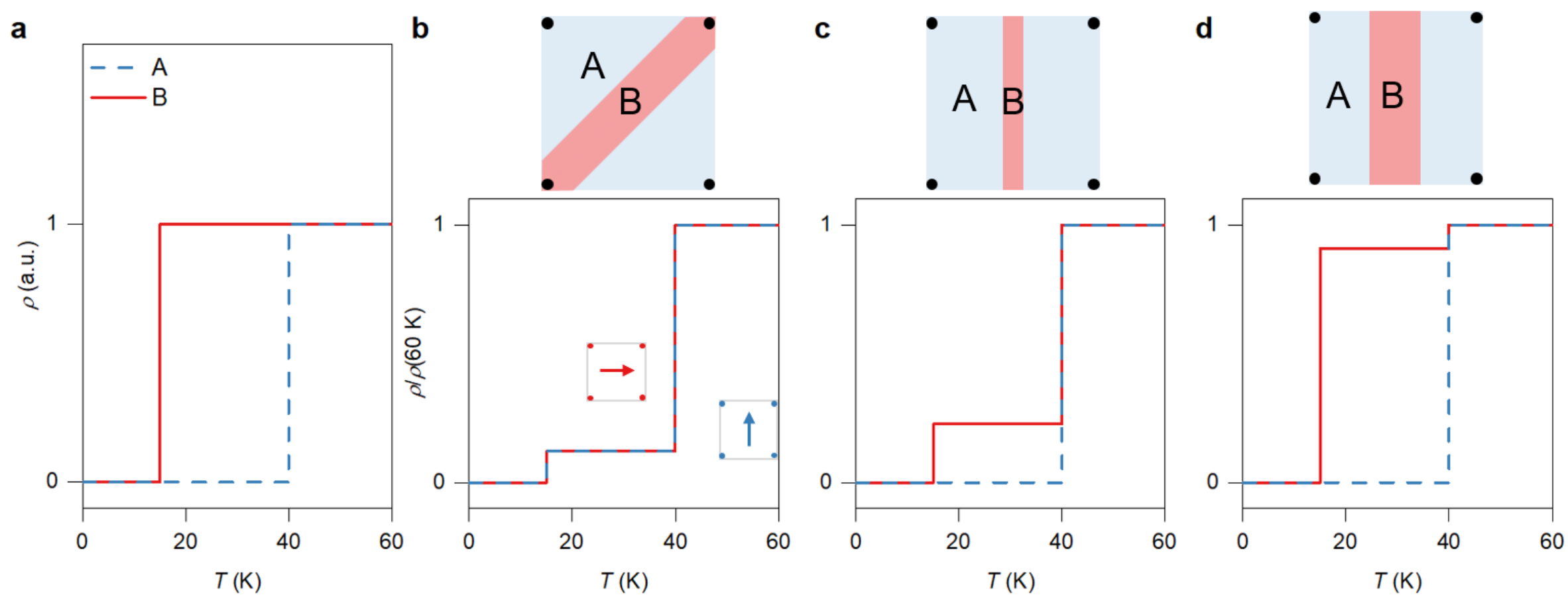

**Figure 3. Simple model illustrating the key ingredients of the two-step superconducting transition and its anisotropy. (a)** Schematic temperature dependence of the resistivity for two distinct regions in the film, labeled A and B, which are assumed to have different $T_c$. **(b-d)** Calculated normalized resistivity $\rho(T)/\rho(60\ \mathrm{K})$ obtained using two van der Pauw configurations for different spatial arrangements of regions A (light blue shaded) and B (light red shaded), as illustrated in the schematics above. Black dots at the four corners of the square denote electrical contacts. In **(b)**, a diagonal arrangement of regions A and B yields identical two-step superconducting transitions in the two measurement configurations (blue and red curves), reproducing the behavior observed in **Figure 1b**,**e**. In **(c)**, a vertically centered B domain gives rise to pronounced anisotropy between the two configurations, similar to that seen in **Figure 1c**,**f**. As the fractional area of domain B increases (or, equivalently, as its absolute resistivity contribution becomes larger), the model in **(d)** reproduces the extreme anisotropic behavior shown in **Figure 1d**,**g**. The current directions for the red solid lines and blue dashed lines are horizontal and vertical, respectively, as illustrated in the inset of **(b)**.

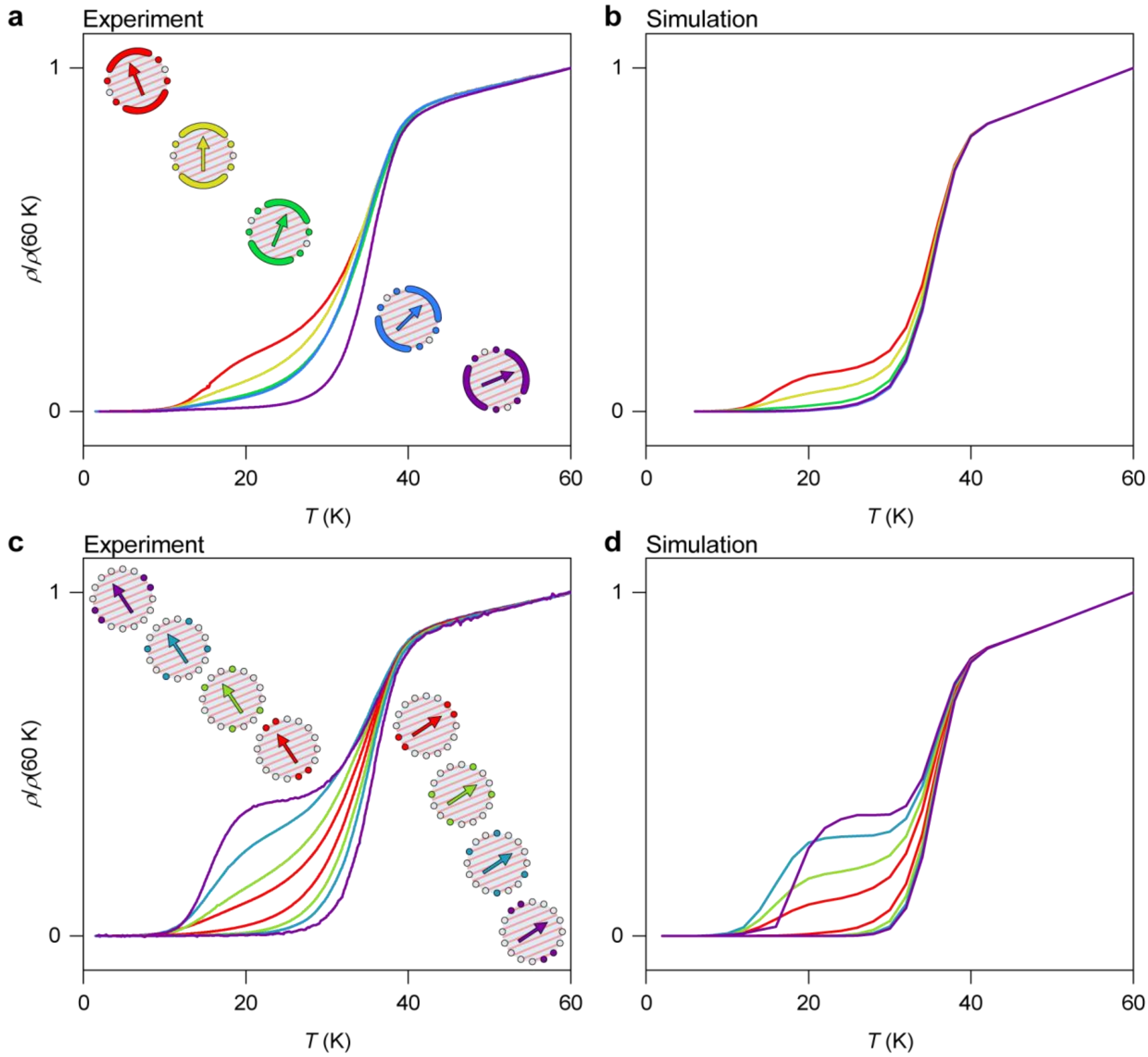

**Figure 4. Finite-element simulation of the superconducting transition. (a,c)** Color-coded temperature-dependent normalized resistivity measured for different in-plane current directions and configurations, as illustrated by the insets. **(b,d)** Finite-element simulations incorporating pristine superconducting regions and edge weak-link regions showing qualitative agreement with the experimentally observed two-step and anisotropic transition behavior.

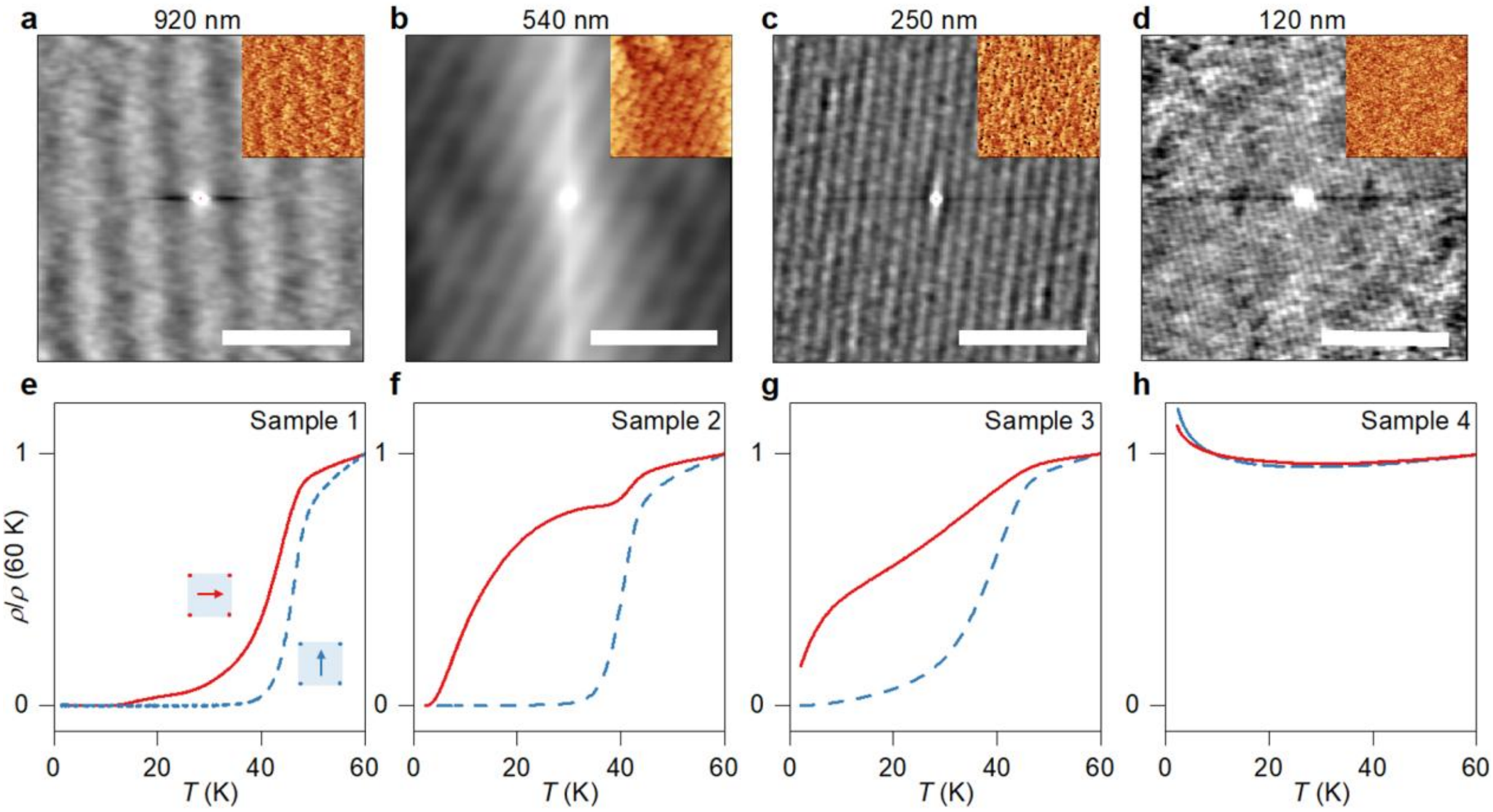

**Figure 5. Influence of step-terrace width on the superconducting transition. (a-d)** Two-dimensional autocorrelation map obtained from AFM topographies of four bilayer nickelate thin-film samples highlighting the width of the terraces, labeled at the top of each map. The direction of the terraces are roughly along the vertical axis. Insets show the corresponding AFM topographies with a 5 × 5 μm field of view, with the contour spanning height ranges of 1 nm for **(a)** and 2 nm for **(b-d)**. We note that the AFM topography in **(d)** was acquired after gentle annealing of the as-grown sample, after which the slightly straightened step edges become resolvable by AFM, primarily to extract a nominal average step-terrace width rather than to capture the intrinsic surface morphology. **(e-h)** Corresponding temperature-dependent normalized resistivity in two van der Pauw configurations for each sample. Solid red curves correspond to measurement configurations in which the current predominantly flows across the step terraces, while dashed blue curves correspond to configurations where the current flows predominantly parallel to the step-terrace direction, as illustrated in the inset of **(e)**. Scale bar in **(a-d)**: 2 μm.

**ToC figure**

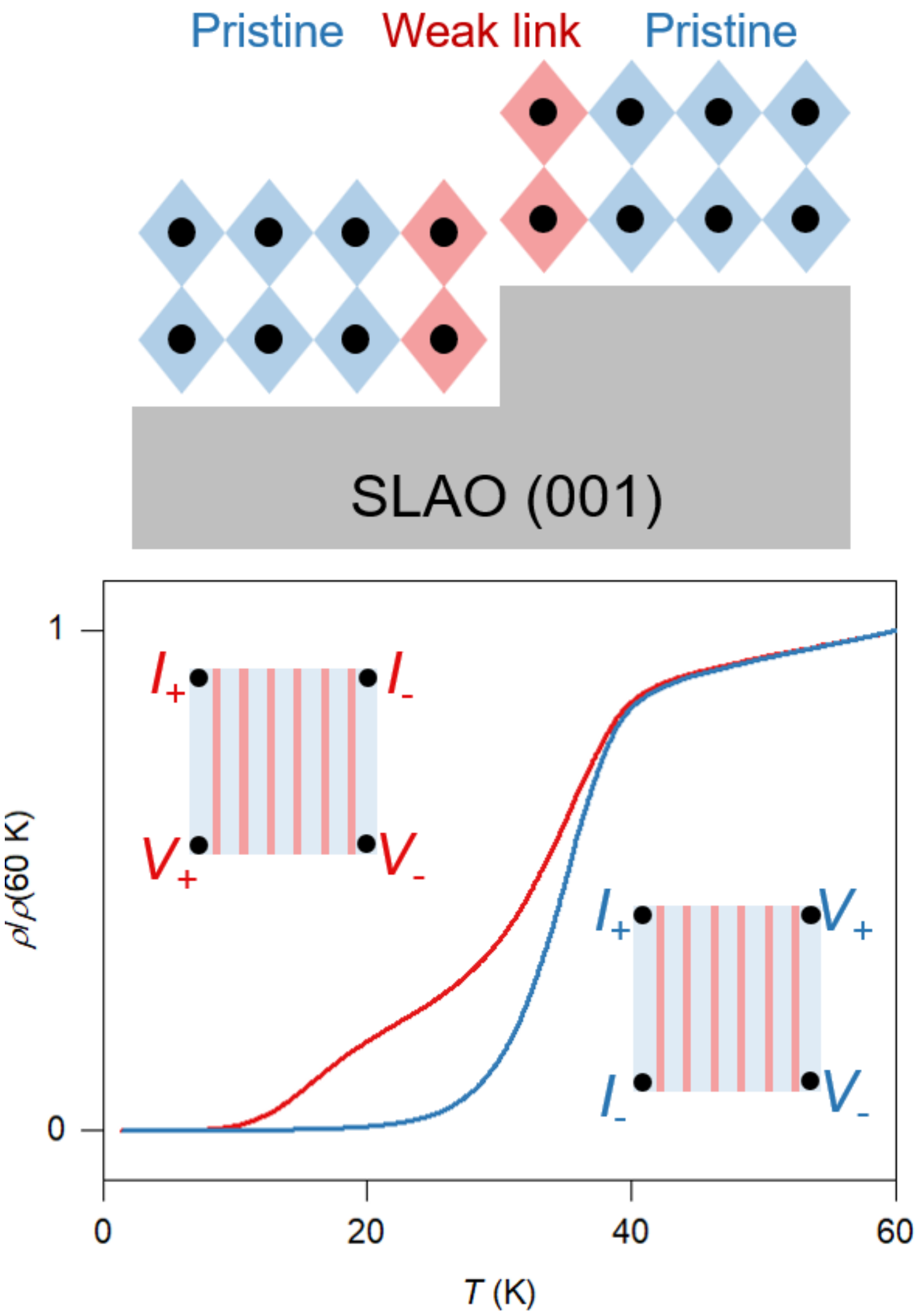

Pristine
Weak link
Pristine
SLAO (001)
$I_+$
$I_-$
$V_+$
$V_-$
$I_+$
$V_+$
$I_-$
$V_-$
1
0
$\rho/\rho$(60 K)
0
20
40
60
$T$ (K)